\documentclass[pra,twocolumn,superscriptaddress,showpacs,preprintnumbers,amsmath,amssymb]{revtex4-1}

\usepackage{graphicx}
\usepackage{dcolumn}
\usepackage{bm}
\usepackage{enumerate}
\usepackage{mathrsfs}
\usepackage{amsmath}
\usepackage{epstopdf}
\usepackage[titletoc]{appendix}
\usepackage[colorlinks=true,breaklinks=true,linkcolor=blue,citecolor=blue,urlcolor=blue]{hyperref}
\usepackage{color}
\usepackage{braket}

\begin{document}
\bibliographystyle{apsrev} 

\title{ Noise-resistant quantum memory enabled by Hamiltonian engineering  }

\author{Lei Jing}
\affiliation{Key Laboratory of Artificial Micro- and Nano-structures of Ministry of Education, School of Physics and Technology, Wuhan University, Wuhan, Hubei 430072, China}

\author{Peng Du}

\affiliation{Key Laboratory of Artificial Micro- and Nano-structures of Ministry of Education, School of Physics and Technology, Wuhan University, Wuhan, Hubei 430072, China}

\author{Hui Tang}
\affiliation{Key Laboratory of Artificial Micro- and Nano-structures of Ministry of Education, School of Physics and Technology, Wuhan University, Wuhan, Hubei 430072, China}

\author{Wenxian Zhang}
\email[Corresponding email: ]{wxzhang@whu.edu.cn}
\affiliation{Key Laboratory of Artificial Micro- and Nano-structures of Ministry of Education, School of Physics and Technology, Wuhan University, Wuhan, Hubei 430072, China}
\affiliation{Wuhan Institute of Quantum Technology, Wuhan, Hubei 430206, China}

\date{\today}
\begin{abstract}

Nuclear spins in quantum dots are promising candidates for fast and scalable quantum memory. By utilizing the hyperfine interaction between the central electron and its surrounding nuclei, quantum information can be transferred to the collective state of the nuclei and be stored for a long time. However, nuclear spin fluctuations in a partially polarized nuclear bath deteriorate the quantum memory fidelity. Here we introduce a noise-resistant protocol to realize fast and high-fidelity quantum memory through Hamiltonian engineering. With analytics and numerics, we show that high-fidelity quantum state transfer between the electron and the nuclear spins is achievable at relatively low nuclear polarizations, due to the strong suppression of nuclear spin noises. For a realistic quantum dot with $10^4$ nuclear spins, a fidelity surpassing 80\% is possible at a polarization as low as 30\%. Our approach reduces the demand for high nuclear polarization, making experimentally realizing quantum memory in quantum dots more feasible.

\end{abstract}

\maketitle

\section{Introduction}
Quantum memory (QM) is a fundamental building block in quantum computation and quantum communication~\cite{briegel1998quantum, wehner2018quantum, sangouard2011quantum, ritter2012elementary, heshami2016quantum}. Although efficient and an-hour-long-storage-time QM has been realized in trapped ions and atomic ensembles~\cite{wang2021single, lvovsky2009optical, choi2010entanglement, julsgaard2004experimental}, fast and scalable solid state candidates for a practical QM are still in demand. Those solid-state physical systems include Nitrogen-vacancy centers in diamonds, doped ions in crystals, and semiconductor quantum dots (QDs)~\cite{bradley2019ten, neumann2010quantum, fuchs2011quantum, ruskuc2022nuclear, hedges2010efficient, morton2008solid, petta2005coherent, koppens2005control, taylor2003long, dobrovitski2006long, ding2014high, gangloff2019quantum, chekhovich2020nuclear, jackson2021quantum, gillard2022harnessing}.
Among them, the QD-based QM which takes the nuclear spin ensemble as its memory medium, is known for its long storage time, excellent optical and electronic properties, and large-area manufacture potentials, making it an appealing option for quantum information processing~\cite{taylor2003long, dobrovitski2006long, ding2014high, gangloff2019quantum, chekhovich2020nuclear, jackson2021quantum, gillard2022harnessing}.

The original QD-based QM protocol is proposed by Taylor, Marcus and Lukin (hereafter referred to as the resonant QM)~\cite{taylor2003long}. It utilizes the hyperfine interaction to write in and read out the quantum information from the electron spin to nuclear spins. Due to the intrinsic long coherence time of nuclear spins, the quantum information can be stored for up to milliseconds~\cite{wust2016role, chekhovich2015suppression, chekhovich2020nuclear, gillard2022harnessing}. In addition, the writing and reading process could be as fast as nanoseconds because of the strong hyperfine coupling between the electron and nuclei. However, the performance of QD-based QM depends sensitively on the nuclear spin polarization. To write an arbitrary qubit into the nuclei with 100\% fidelity, full polarization is required, which is impossible to achieve in practice. Recent advances in optical pumping of nuclear spins in GaAs/AlGaAs QDs have illustrated about 80\% nuclear polarization~\cite{chekhovich2017measurement}. Even with such a record-high degree of polarization, the fidelity of the resonant QM protocol is still below 80\%~\cite{taylor2003long, dobrovitski2006long}. Alternative approaches to ensure high QM fidelity but at low nuclear polarizations are constantly in great demand, such as nuclear state preparation~\cite{ethier2017improving, reilly2008suppressing, gangloff2019quantum, evers2021suppression}, inhomogeneous polarization~\cite{wu2016inhomogeneous, ding2014high},
 and the use of noncollinear hyperfine interaction~\cite{denning2019collective, jackson2021quantum}.

The major obstacle in QD-based QM protocols stems from the nuclear spin fluctuations, which become prominent at low polarizations and degrade significantly the QM fidelity. To suppress the nuclear spin noises, we propose in this paper a noise-resistant QM (NRQM) protocol through Hamiltonian engineering of electron-nuclear hyperfine interaction. By applying periodically fast $\pi$-pulses along $x$- and $y$-axis, the hyperfine interaction is effectively transformed into a flip-flop Hamiltonian, which simultaneously flips the electron spin and flops a nuclear spin thus realizes an efficient quantum state transfer. More importantly, the effects of nuclear spin fluctuations are strongly suppressed by these pulses. With this idea, we realize high-fidelity QM but at relatively low polarizations. Our scheme is compatible with inhomogeneous polarization and nuclear state preparation. Better QM performance is achieved by combining them together.

\section{Electron-nuclear spin dynamics in a QD-based QM}
For QDs, the coupling of the $s$-state conduction electron to a mesoscopic bath of nuclear spins is governed by hyperfine contact interaction~\cite{khaetskii2002electron, merkulov2002electron, philippopoulos2020first}.
In a magnetic field $B_0$ along the $z$-axis, the Hamiltonian is
\begin{equation}
 H = g_e^* \mu_B B_0  S^z + \sum_{j} A_j \mathbf{I}_j  \cdot  \mathbf{S} ,
 \label{eq:H0}
 \end{equation}
where the first term corresponds the electron Zeeman energy. Spin operators $\mathbf{S}$ and $\mathbf{I}_j$ are for the electron and the $j$-th nucleus, respectively. We assume all spins are spin-1/2 for convenience. The coupling strength $A_j$  is given by $A_j = A_0 v_0 \left| \psi(\mathbf{r}_j) \right|^2$ with $A_0$ being the hyperfine contact interaction constant, $v_0$ the volume of a unit cell and $\left| \psi(\mathbf{r}_j) \right|^2$ the probability density of the electron at site $\mathbf{r}_j$ of the $j$-th nucleus~\cite{slichter2013principles}.
 The $A_j$ is a function of position varying in a Gaussian form in a typical QD~\cite{johnson2005triplet, koppens2005control, petta2005coherent}.
 The hyperfine interaction term can be rewritten as $ H_D + H_\Omega$, where $H_D =  S^z \sum_j A_j I_j^z$ and $ H_\Omega = \frac{1}{2}\sum_{j} A_j \left( S^+ I^-_j + S^- I^+_j \right) $ with $S^{\pm} = S^x \pm i S^y$ and $I_j^{\pm} = I_j^x \pm i I_j^y$. The diagonal term $H_D$ produces an effective magnetic field on the electron called the Overhauser field. By tuning the magnetic field $B_0$ to be equal in magnitude and opposite to the Overhauser field, $H_D$ may be cancelled and only the flip-flop term $H_\Omega$ is left, which introduces spin exchange between the electron and the nuclei.

Utilizing the flip-flop term $H_{\Omega}$ and zeroing the sum of the Zeeman term and $H_D$, Taylor, Marcus and Lukin proposed a resonant QM protocol in a QD~\cite{taylor2003long}. Starting with a fully polarized nuclear bath
$\ket{\mathbf{0}}_n = \ket{I_0, \dots, I_0}_n$
and a spin-down electron $\ket{\downarrow}_e$, this flip-flop Hamiltonian $H_\Omega$ induces a flipped electron spin with a collective nuclear spin excitation $\ket{\uparrow}_e \bigotimes \ket{\mathbf1}_n $ where
$\ket{\mathbf{1}}_n = \left( \sum_j \left| A_j \right|^2 \right)^{-1/2} \sum_j A_j \ket{I_0, \dots, (I_0 - 1)_{(j)}, \dots, I_0}_n$.
Since a spin-up electron and a fully polarized bath stay still due to the conservation of angular momentum, an arbitrary initial electron spin evolves like
\begin{equation}
\left( \alpha \ket{\uparrow}_e + \beta\ket{\downarrow}_e \right) \otimes \ket{\mathbf{0}}_n \to  \ket{\uparrow}_e \otimes \left( \alpha \ket{\mathbf{0}}_n + i\beta \ket{\mathbf{1}}_n \right).
\label{eq:1}
\end{equation}
In this way the quantum state of the electron spin is coherently mapped into the collective mode of the nuclei. The quantum state transition can be turned off by removing the electron from the QD, tuning the magnetic field away from the resonant condition or dynamical decoupling. Due to nuclear spins' long coherence time, information encoded in the nuclear spins can be preserved for a long time~\cite{chekhovich2015suppression, chekhovich2020nuclear}. Retrieval of the stored information is simply reversing the process: let the system oscillate for another half cycle under the flip-flop Hamiltonian and the quantum information returns to the electron.

In practice, full nuclear polarization is difficult to achieve. Incomplete polarization may degrade significantly the QM performance. For instance, a partly polarized thermal nuclear bath is composed of  many different pure nuclear spin states $\ket{I, M}$, $\rho = \sum w(I, M) \ket{I, M} \bra{I, M} $, where $I$ is the total angular momentum for $N$ nuclear spins and $M = -I, -I+1,\cdots, I$ is its projection into the $z$-axis~\cite{slichter2013principles}.
To transfer the qubit back and forth fully between the electron and the nuclear state, all pure states have to be simultaneously in resonance. This is roughly the case at high polarization. However, at low polarization $P$, the probability of $I$ follows approximately a Gaussian distribution with a width $\sigma = \sqrt{(1-P)(1+P)N/4}$ increasing as $P$ decreases, indicating that a large number of bath states are off-resonant~\cite{taylor2003controlling}. These off-resonant oscillations damp the Rabi oscillation and deteriorate the QM performance~\cite{taylor2003long, taylor2003controlling}.

\begin{figure}[]
  \includegraphics[width=3.40in]{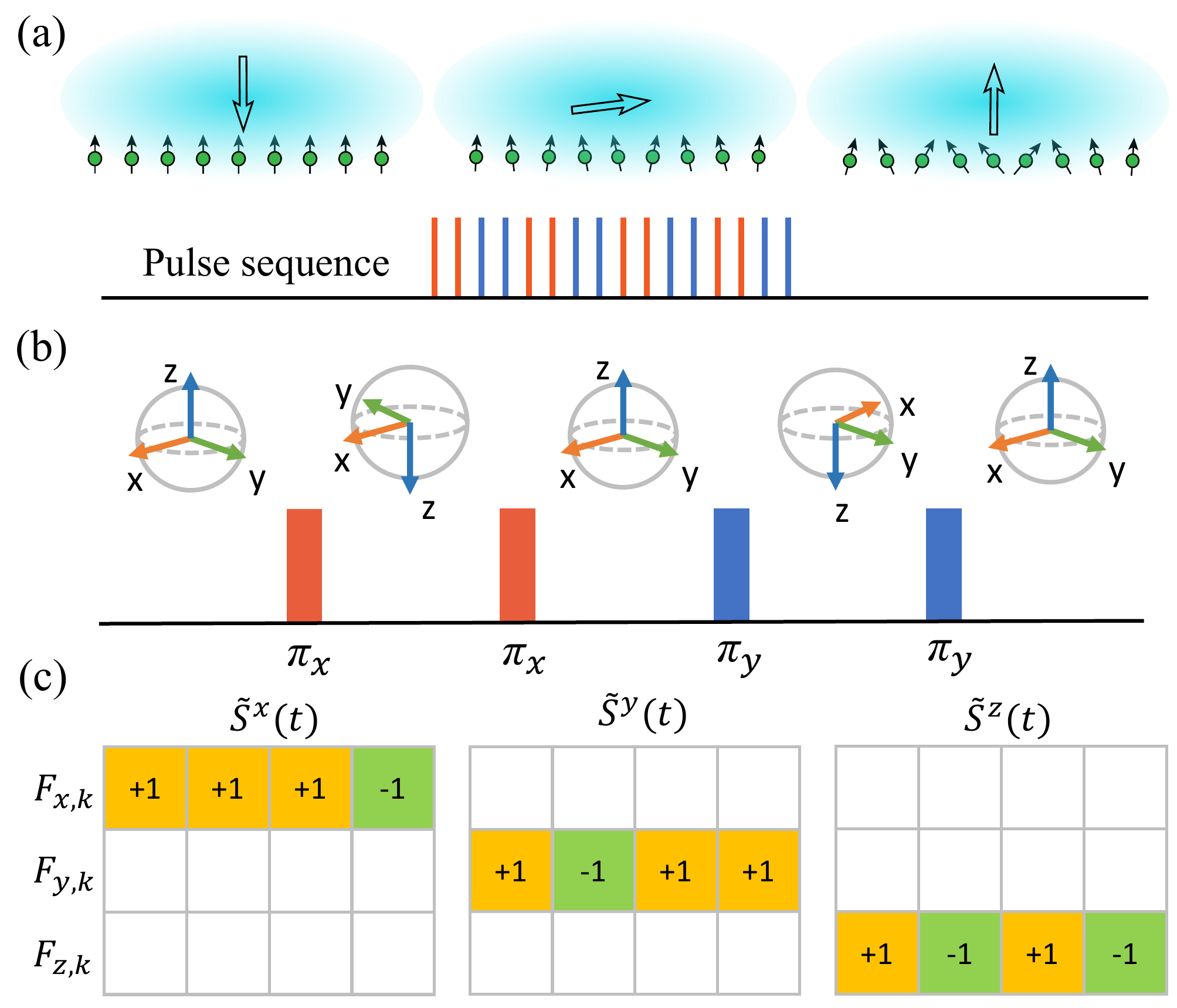}
  \caption{
  (a) Procedure of NRQM, where the electron spin state is coherently transferred into the collective modes of nuclei through the flip-flop Hamiltonian $H_{\Omega}$ enabled by pulses.
  (b) A pulse cycle [XXYY] for Hamiltonian engineering to generate effectively $H_{\Omega}$. The spin frames (spin operators instead of states) shown as spheres are periodically rotated by the pulses in the interaction picture.
  (c) Time-domain transformations of spin operators $\tilde{S}^{x, y, z}(t) $ in the interaction frame driven by the periodic pulse sequence, depicted by the matrix-based representation $[F_{\mu,k}]$, where the row is $\mu = (x, y, z)$ and the column is $k = (1, 2, 3, 4)$.
    }
  \label{fig:figure1}
\end{figure}

\section{Noise-resistant quantum state transfer via Hamiltonian engineering}
To implement the QM in a QD even at low nuclear polarizations, we need to design a  pulse sequence that keeps intact the desired flip-flop Hamiltonian but significantly suppresses the fluctuation of the Overhauser field, which causes the off-resonant oscillations. The developed pulse sequence is $\text{[XXYY]}^n$, $n$ cycles of [XXYY] as shown in Fig.~\ref{fig:figure1}, where X and Y represent a $\pi$ pulse that rotates the electron spin 180 degrees around the $x$- and $y$-axis, respectively. The pulse interval is $\tau$.

It is straightforward to illustrate the effect of the pulse sequence according to the average Hamiltonian theory and the matrix representation~\cite{choi2020robust}.
In the toggling frame the electron spin operators are periodically rotated by the pulses, leading to the time-dependent Hamiltonian
\begin{equation}
\tilde{H}(t) =  \sum_j \left[A_j  I^x_j \tilde{S}^x(t) \!+\! A_j I^y_j \tilde{S}^y (t) \right]\!+\! g_e^* \mu_B B_{\text{eff}}\tilde{S}^z (t) ,
\end{equation}
where $B_{\text{eff}} = B_0 + \sum_j A_j I_j^z /   g_e^* \mu_B$ is the effective magnetic field. Taking the spin operator $S^z$ as an example, the pulse sequence transforms it into $\pm S^z$ operators periodically. For each spin operator $S^i$, we can identify its transformation trajectory as
\begin{equation}
\tilde{S}^i (t) = \sum_\mu F_{\mu, k} S^\mu,  \,\,\,\,\,\,  \text{for} \,\, t_{k-1} < t < t_k.
\end{equation}
where the $\mathbf{F} = [F_{\mu, k}] = [F_{x, k}; F_{y, k}; F_{z, k}]$ is a $3 \times n$ matrix containing only $0$ and $\pm 1$, $t_k$ is the time point at which the pulses are applied.
As depicted in Fig.~\ref{fig:figure1}(c), the transformation matrix representation for $\tilde S^i$ ($i={x,y,z}$) reveals how the pulses alter the system's spin dynamics in an intuitive way. By averaging $\tilde S^i(t)$, we find that the ${S}^z$ operator is effectively cancelled but a fraction of  ${S}^x$ and ${S}^y$ remains. Thus the pulse sequence zeros the effective magnetic field while maintaining the flip-flop interaction. In this way, one easily obtains the  zeroth-order average Hamiltonian
\begin{equation}
\overline{H}^{(0)} = \frac{1}{4} \sum_j A_j  \left( S^+I^-_j + S^-I^+_j \right) .
\label{eq:average_H}
\end{equation}
Higher-order terms are neglected since they diminish as $\tau$ approaches zero.

The above analysis indicates that the pulse sequence  [XXYY]$^n$ indeed generates the desired flip-flop Hamiltonian. In addition, this protocol is expected to be robust against magnetic noise in the $z$-direction (e.g. fluctuations of the Overhauser field) because terms containing $S_z$ in the Hamiltonian average to 0. In this sense, we refer to the protocol as the NRQM. Compared to the resonant QM protocol, NRQM is independent of the external magnetic field $B_0$ and may outperform the resonant one, particularly at lower nuclear polarizations.

\section{Analytical results for $A_j=A$}The nuclear bath is composed of $10^4$ to $10^6$ nuclear spins, each having different coupling strength $A_j$ with the electron. Analytics for the dynamics of the system is challenging~\cite{coish2004hyperfine, bortz2007exact, gaudin1976diagonalisation, faribault2013integrability, schliemann2003electron}. In the following we consider the case where inhomogeneity is negligible ($A_j = A$), such that the dynamics of the system under the average Hamiltonian $\overline{H}^{(0)}$ is analytically solvable, and we analyze the performance of the NRQM protocol following the derivation in Ref.~\cite{dobrovitski2006long}.

For the initial electron state  $\ket{\uparrow} $ and the collective nuclear state $\ket{ I, M }$, the system's wave function after time $t$ is
$ \ket{\psi_1(t) } = \cos(\omega_1 t) \ket{\uparrow} \otimes \ket{ I, M }     -     i \sin (\omega_1 t) \ket{\downarrow} \otimes \ket{ I, M + 1 }$ where $\omega_1 = A \sqrt{ (I - M)(I + M + 1)} / 4 $. For the initial electron state $\ket{\downarrow} $ the system evolves as $\ket{ \psi_2(t) } = \cos(\omega_2 t) \ket{ \downarrow } \otimes \ket{ I, M }     -     i \sin (\omega_2 t) \ket{\uparrow} \otimes \ket{ I, M - 1 } $, where $\omega_2 = A \sqrt{ (I + M)(I - M + 1)} / 4 $.
Obviously, the oscillation frequencies of the dynamics depends on $I$ and $M$.

For a partially polarized nuclear bath in thermal equilibrium, the statistical weight $w(M) = C^N_{k} \theta^{k} (1-\theta)^{k}$, where $C^N_{k}$ is the binomial coefficient, $k = N/2-M$ and  $\theta = e^\gamma / (1 + e^\gamma)$~\cite{taylor2003controlling}. The corresponding nuclear polarization is $P=\tanh (\gamma / 2)$. This distribution is roughly Gaussian centered at $\overline{M} = -NP / 2$ with the variance $\sigma^2 = (N/4)(1-P)(1+P)$. The statistical weight of the state $\ket{ I, M } $ is
\begin{equation}
\begin{aligned}
w(I, M) &= w(M)(C_{m}^N - C_{m - 1}^N) / C_{k}^N, \\
 & \approx w(M) \zeta ^{I - |M|},
 \end{aligned}
 \label{eq:approx}
\end{equation}
where $ m = N/2 - I$, $\zeta =  \Delta P/(2 - \Delta P)$ and $\Delta P = 1 - P  $. The approximation in Eq.~(\ref{eq:approx}) holds for small ${I - |M|}$. One immediately finds that for a large value of $P$ (small $\Delta P$ and $\zeta$), the states $\ket{ I, M } $ with $M = -I$ account for the majority with a proportion of $1-\zeta$.
Other states with higher $I (> -M)$ have an exponentially smaller statistical weight and have much less impact on the dynamics.
To calculate the fidelity of the NRQM up to the linear order in $\zeta$, we only include the states with $ M = -I$ and $ M = - I + 1$ and neglect the others.

\begin{figure}
  \centering
  \includegraphics[width=3.35in]{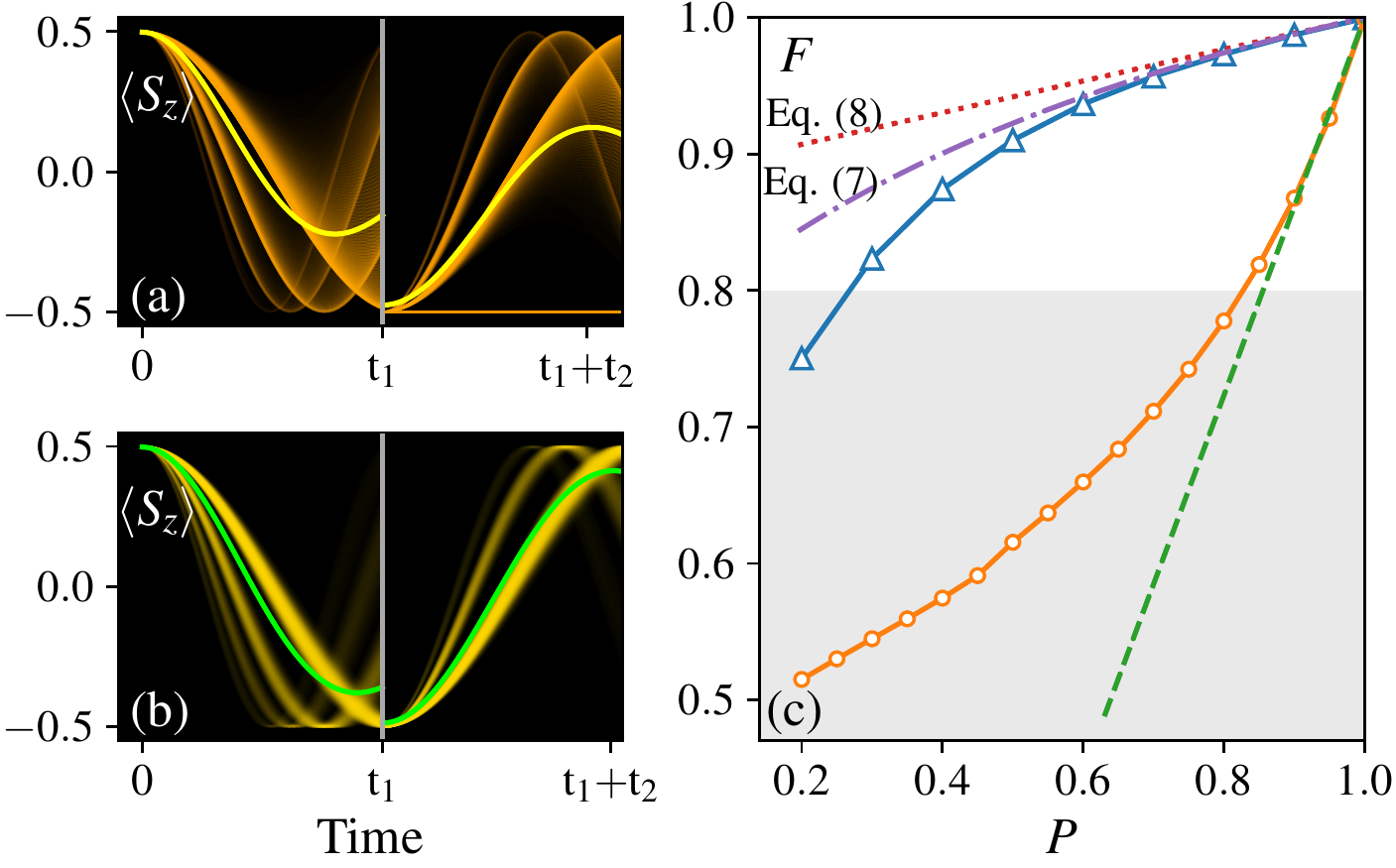}
  \caption{ Typical dynamics of the electron spin in a partially polarized nuclear spin bath ($P=0.5$) (a) in the resonant QM protocol and (b) in the NRQM protocol.  Because the bath is a mixture of numerous potential states, the system's evolution can take many different paths. When we depict every path, we get blurred lines in the image that diffuse over time.
  The ensemble averaged dynamics of the electron spin shown as the yellow and green solid lines are damped Rabi oscillations. (c) Fidelity as a function of nuclear bath polarization $P$ with $10^4$ nuclear spins for the resonant QM protocol (solid line with circles) and NRQM protocol (solid line with triangles). Dash-dotted and dotted lines: analytical estimates from Eqs.~(\ref{eq:F_Delta}) and (\ref{eq:F_Delta_P}), respectively, for the NRQM protocol. Dashed line: analytical estimate for the resonant QM protocol~\cite{dobrovitski2006long}. The NRQM protocol significantly outperforms the resonant QM.
 }
  \label{fig:figure2}
\end{figure}

The fidelity of the memory protocol is defined as the overlap between the initial electron state and the retrieved one, taking the minimum over all possible initial states. According to Ref.~\cite{dobrovitski2006long}, the minimal fidelity can be found by considering only two types of initial conditions:
(1) the electron spin pointing to the $z$-axis, (2) the electron spin lying in the $xy$-plane. For the first case, we calculate $s_z = \mathrm{Tr}(S^z \rho_f)$, where $\rho_f$ is the density matrix for the final retrieved electron state. For the second case,  we calculate  $s_T = \sqrt{s_x^2 + s_y^2}$ and $s_0 = \mathrm{Tr}(S^z \rho_f)$, where  $s_{x, y} = \mathrm{Tr}(S^{x, y} \rho_f)$. The fidelity $F$ is the minimum of the following three quantities:
$   f_1 = (1 +s_z)/2$,
$   f_2 = (1 + s_z - 2s_0)/2$ and
$   f_3 = ( 1 + s_T - s_0^2/[4(s_z - s_0 - s_T)]) / 2
$,
When $ s_0/[2(s_z - s_0 - s_T)] \notin [-1, 1]$, $F$ takes the minimum of $f_1$ and $f_2$.

By taking into account of two types of initial bath states, $M = -I $ and $ M = - I +1$, and averaging with probability $w(I, M)$, we calculate $s_z$, $s_0$, $s_T$ up to the linear order in  $\zeta$ :
$
s_z  \approx 1 - \zeta \left( 1  - \cos^4\gamma_0 - \sin^4 \gamma_0   \right)   ,
s_0 \approx - ({\zeta}/{2}) \left(    1  - \cos^4\gamma_0 - \sin^4 \gamma_0   \right) ,
s_T \approx 1 - \zeta \left(    1 + \cos \gamma_0    \right)   ,
$
 where $\gamma_0 = \pi / \sqrt{2}$.We then obtain the fidelity
 \begin{equation}
 \begin{aligned}
 F =\frac {1+s_z} 2 &\approx 1 - 0.232\, \zeta \; .
 \label{eq:F_Delta}
 \end{aligned}
 \end{equation}
Substituting $\zeta= \Delta P / (2-\Delta P)$, the fidelity is further simplified as
\begin{equation}
F \approx 1 - 0.116 \Delta P\; .
\label{eq:F_Delta_P}
\end{equation}

The results from Eqs.~(\ref{eq:F_Delta}) and (\ref{eq:F_Delta_P}) are depicted in Fig.~\ref{fig:figure2}(c) as dash-dotted and dotted lines.
For comparison, the analytical estimate of fidelity for the original resonant QM protocol is $F \approx 1 - 1.38 \Delta P $~\cite{dobrovitski2006long}, which is also depicted in Fig.~\ref{fig:figure2}(c) as the dashed line.
Starting with full polarization, the fidelity of the NRQM protocol drops more than ten times slower than that of the resonant one as the nuclear polarization lowers, demonstrating the advantages of the NRQM protocol.

\section{Numerical results}
While we have derived the fidelity Eq.~(\ref{eq:F_Delta}) at high bath polarizations and proved the advantage of the NRQM, the performance of the protocol at low polarizations or inhomogeneous hyperfine coupling is still in need. This task is completed by numerical simulations.

First, we consider the case where inhomogeneity is negligible ($A_j = A$). We numerically simulate the dynamics of the electron spin and $N=10^4$ nuclear spins.
The system is prepared as a tensor product of the electron state $\ket{\phi}$ and the nuclear state $\ket{I, M}$ with a statistical weight $w(I, M)$.
The system evolves under the Hamiltonian Eq.~(\ref{eq:H0}) and the pulse sequence [XXYY]$^n$ for a time $t_1$ and the electron state is mapped onto the collective nuclear state. After that, the electron is ejected and a nuclear mixed state is reduced by tracing out the electron's degree of freedom. For the retrieval, another fully polarized electron is injected in the QD. The system's state becomes a tensor product of the electron state and the reduced nuclear bath state. Then the system evolves under the same pulse sequence for another time $t_2$. After tracing out the nuclear bath's degree of freedom, we obtain the final density matrix $\rho_f$ of the electron.
We plot how the electron observable $\braket{S_z}$ changes over time during this process in Fig.~\ref{fig:figure2}(a) and (b).
For a given $|I,M\rangle$, the trajectory of $\braket{S_z}$ is a translucent curve with  opacity based on the statistical weight of the bath state $w(I,M)$.
In total, we get blurred lines in the image that diffuse over time.
The thickest blurred line, which in fact includes many lines, shows the electron spin's evolution with bath states $\ket{I, M}$ where $M = -I$, $I \in [0, N/2]$. The rest blurred lines come from bath states $\ket{I, M}$ where $M = -I + 1$, $M = -I + 2$, etc.
By averaging over different bath states with statistical weight $w(I, M)$ we calculate the weighted observables for the electron ($s_{z,0,T}$) in a realistic thermal bath ensemble.
This ensemble averaged dynamics of the electron are drawn as the yellow and green solid lines in Fig.~\ref{fig:figure2}(a) and (b).
The minimal fidelity is calculated accordingly using ($s_{z,0,T}$)~\cite{dobrovitski2006long}. The times $t_1$ and $t_2$ are tuned to maximize the minimal fidelity. The NRQM results at different nuclear polarization $P$ are shown in Fig.~\ref{fig:figure2}(c) as a solid line with triangles. The simulated fidelities for the resonant QM protocol are also plotted as a solid line with circles for comparison.

As shown in the figure, the NRQM illustrates a significant improvement in the fidelity over the resonant QM.
The NRQM protocol has a fidelity over $90\%$ at $P = 0.5$ and the fidelity is still over $80\%$ at $P = 0.3$.
In stark contrast, the resonant QM protocol requires a polarization greater than $0.8$ in order to achieve $80\%$ fidelity~\cite{taylor2003long, dobrovitski2006long}. Clearly, the need for strong nuclear polarization is dramatically mitigated for the NRQM protocol.
In addition, the numerics agrees well with our analytical estimate of the fidelity at high polarizations, implying the validity of previous analytics.

\begin{figure}
  \includegraphics[width=3.35in]{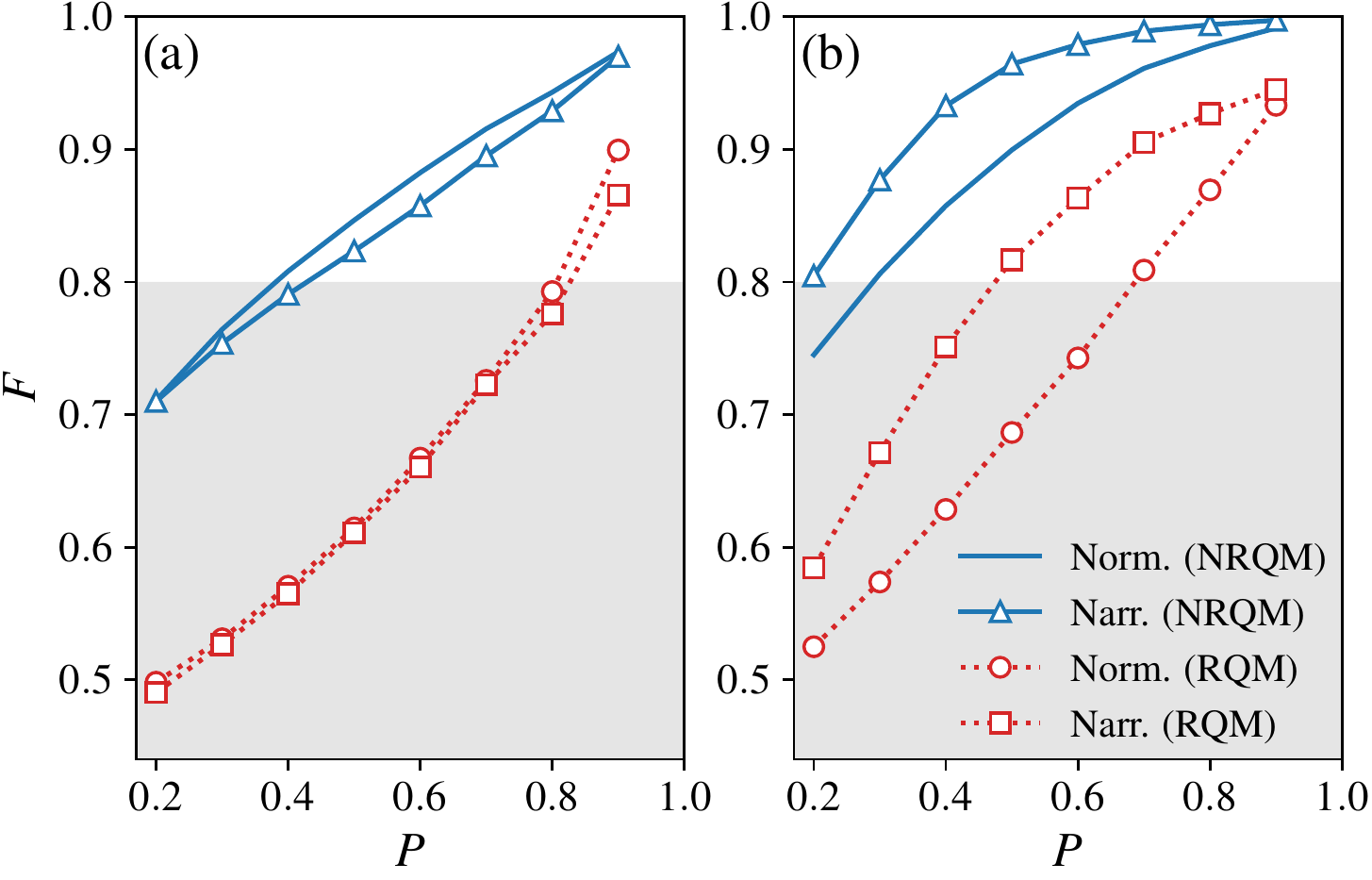}
  \caption{Performance of quantum memory protocols in a bath of N=20 nuclear spins. (a) QM fidelity for homogeneous bath polarizations with different widths of $A_j$ for the resonant QM protocol (RQM) and the NRQM protocol. (b) QM fidelity for inhomogeneous bath polarizations.
  }
  \label{fig:figure3}
\end{figure}

Second, we consider the case of inhomogeneous hyperfine coupling ($A_j \neq A$). We numerically simulate the dynamics of the electron spin interacting with $ N=4\times 5$ nuclear spins arrayed in a rectangular lattice, using the efficient Chebyshev-expansion-based algorithm~\cite{dobrovitski2003efficient}.
More computer resources would be required to include more nuclear spins, but $N=20$ is adequate to make our simulations represent bigger systems with a precision of $1/N$~\cite{dobrovitski2006long}.
 In order to compare with previous works, we adopt the same coupling strengths $A_j$ as in Refs.~\cite{dobrovitski2006long, ding2014high}.
The values of $A_j$  spread from 0.96 to 0.31, referred to as the normal distribution. We also consider the case with decreased QD widths by a factor of $1/\sqrt{2}$ to represent a narrow distribution of $A_j$, spreading from 0.92 to 0.09.

Numerical results are presented in Fig.~\ref{fig:figure3}(a). As shown in the figure, the NRQM outperforms significantly the resonant QM. The fidelity jumps from around 60\% up to 80\% at $P = 0.5$ if one switches from the resonant QM to the NRQM. In addition, normal and narrow distribution of $A_j$ have little impact on the fidelity. Thus the results for $A_j=A$ are expected similar to that for $A_j \neq A$. Such an independence of the distribution of $A_j$ indicates that the results shown in Fig.~\ref{fig:figure2}(c) may also be applicable to the case $A_j \neq A$ even for $N=10^4$. Therefore, we expect that the NRQM performs better than the resonant QM in a realistic QD.

Finally, we look at the case of inhomogeneous nuclear polarization, which can be produced by dynamic nuclear polarization (DNP)~\cite{reilly2008suppressing, chekhovich2017measurement}.
In DNP, the speed of an individual nuclear spin's polarization is roughly proportional to the square of its hyperfine coupling strength, resulting in a spatially nonuniform distribution of nuclear polarization in the QD. High degree of polarization occurs at these strongly coupled nuclei.
Polarization of the $j$-th nuclear spin after DNP is approximately $p_j = \tanh (\beta A_j^2)$, where $\beta$ is a parameter related to the number and duration of DNP cycles~\cite{wu2016inhomogeneous, zhang2010protection}.
It was reported that inhomogeneous nuclear polarization significantly improves the performance of a QD-based quantum memory~\cite{ding2014high}.
The performance improves even more when NRQM is used to suppress nuclear spin noise.
Numerical simulations are carried out in the same model with the coupling strength $A_j$ unchanged.
As shown in Fig.~\ref{fig:figure3}(b), the NRQM protocol again outperforms the resonant one for inhomogeneous nuclear polarization, similar to the homogeneous case. However, the performance of the NRQM for the inhomogeneous polarization depends on the distribution of $A_j$, which is quite different from that for the homogeneous case.
In circumstances of the ``narrow" distribution of $A_j$, the combined scheme shows the best performance: a fidelity over 95\% at a bath polarization $P=0.5$, and a fidelity over 80\% at $P=0.2$.

\section{Conclusion}
We proposed  a noise-resistant pulsed quantum memory protocol that performs coherent state transfer between the electronic  and nuclear spins using Hamiltonian engineering of the hyperfine interaction.
Because of its strong suppression of nuclear spin noise, the NRQM protocol reduces the requirement for high nuclear polarization, making experimental realization of QD-based quantum memory more feasible.
In addition, this Hamiltonian engineering approach may be helpful for further investigations in quantum memory and DNP in other systems such as NV color centers, doped-ion crystals and atomic ensembles.

\begin{acknowledgments}
This work is supported by the NSAF under Grant No. U1930201, National Natural Science Foundation of China (NSFC) under Grants No. 12274331 and No. 91836101, and Innovation Program for Quantum Science and Technology under Grant No. 2021ZD0302100. The numerical calculations in this paper have been partially done on the supercomputing system in the Supercomputing Center of Wuhan University.
\end{acknowledgments}


\begin{thebibliography}{46}
\expandafter\ifx\csname natexlab\endcsname\relax\def\natexlab#1{#1}\fi
\expandafter\ifx\csname bibnamefont\endcsname\relax
  \def\bibnamefont#1{#1}\fi
\expandafter\ifx\csname bibfnamefont\endcsname\relax
  \def\bibfnamefont#1{#1}\fi
\expandafter\ifx\csname citenamefont\endcsname\relax
  \def\citenamefont#1{#1}\fi
\expandafter\ifx\csname url\endcsname\relax
  \def\url#1{\texttt{#1}}\fi
\expandafter\ifx\csname urlprefix\endcsname\relax\def\urlprefix{URL }\fi
\providecommand{\bibinfo}[2]{#2}
\providecommand{\eprint}[2][]{\url{#2}}

\bibitem[{\citenamefont{Briegel et~al.}(1998)\citenamefont{Briegel, D{\"u}r,
  Cirac, and Zoller}}]{briegel1998quantum}
\bibinfo{author}{\bibfnamefont{H.-J.} \bibnamefont{Briegel}},
  \bibinfo{author}{\bibfnamefont{W.}~\bibnamefont{D{\"u}r}},
  \bibinfo{author}{\bibfnamefont{J.~I.} \bibnamefont{Cirac}}, \bibnamefont{and}
  \bibinfo{author}{\bibfnamefont{P.}~\bibnamefont{Zoller}},
  \bibinfo{journal}{Phys. Rev. Lett.} \textbf{\bibinfo{volume}{81}},
  \bibinfo{pages}{5932} (\bibinfo{year}{1998}).

\bibitem[{\citenamefont{Wehner et~al.}(2018)\citenamefont{Wehner, Elkouss, and
  Hanson}}]{wehner2018quantum}
\bibinfo{author}{\bibfnamefont{S.}~\bibnamefont{Wehner}},
  \bibinfo{author}{\bibfnamefont{D.}~\bibnamefont{Elkouss}}, \bibnamefont{and}
  \bibinfo{author}{\bibfnamefont{R.}~\bibnamefont{Hanson}},
  \bibinfo{journal}{Science} \textbf{\bibinfo{volume}{362}},
  \bibinfo{pages}{eaam9288} (\bibinfo{year}{2018}).

\bibitem[{\citenamefont{Sangouard et~al.}(2011)\citenamefont{Sangouard, Simon,
  De~Riedmatten, and Gisin}}]{sangouard2011quantum}
\bibinfo{author}{\bibfnamefont{N.}~\bibnamefont{Sangouard}},
  \bibinfo{author}{\bibfnamefont{C.}~\bibnamefont{Simon}},
  \bibinfo{author}{\bibfnamefont{H.}~\bibnamefont{De~Riedmatten}},
  \bibnamefont{and} \bibinfo{author}{\bibfnamefont{N.}~\bibnamefont{Gisin}},
  \bibinfo{journal}{Rev. Mod. Phys.} \textbf{\bibinfo{volume}{83}},
  \bibinfo{pages}{33} (\bibinfo{year}{2011}).

\bibitem[{\citenamefont{Ritter et~al.}(2012)\citenamefont{Ritter, N{\"o}lleke,
  Hahn, Reiserer, Neuzner, Uphoff, M{\"u}cke, Figueroa, Bochmann, and
  Rempe}}]{ritter2012elementary}
\bibinfo{author}{\bibfnamefont{S.}~\bibnamefont{Ritter}},
  \bibinfo{author}{\bibfnamefont{C.}~\bibnamefont{N{\"o}lleke}},
  \bibinfo{author}{\bibfnamefont{C.}~\bibnamefont{Hahn}},
  \bibinfo{author}{\bibfnamefont{A.}~\bibnamefont{Reiserer}},
  \bibinfo{author}{\bibfnamefont{A.}~\bibnamefont{Neuzner}},
  \bibinfo{author}{\bibfnamefont{M.}~\bibnamefont{Uphoff}},
  \bibinfo{author}{\bibfnamefont{M.}~\bibnamefont{M{\"u}cke}},
  \bibinfo{author}{\bibfnamefont{E.}~\bibnamefont{Figueroa}},
  \bibinfo{author}{\bibfnamefont{J.}~\bibnamefont{Bochmann}}, \bibnamefont{and}
  \bibinfo{author}{\bibfnamefont{G.}~\bibnamefont{Rempe}},
  \bibinfo{journal}{Nature (London)} \textbf{\bibinfo{volume}{484}},
  \bibinfo{pages}{195} (\bibinfo{year}{2012}).

\bibitem[{\citenamefont{Heshami et~al.}(2016)\citenamefont{Heshami, England,
  Humphreys, Bustard, Acosta, Nunn, and Sussman}}]{heshami2016quantum}
\bibinfo{author}{\bibfnamefont{K.}~\bibnamefont{Heshami}},
  \bibinfo{author}{\bibfnamefont{D.~G.} \bibnamefont{England}},
  \bibinfo{author}{\bibfnamefont{P.~C.} \bibnamefont{Humphreys}},
  \bibinfo{author}{\bibfnamefont{P.~J.} \bibnamefont{Bustard}},
  \bibinfo{author}{\bibfnamefont{V.~M.} \bibnamefont{Acosta}},
  \bibinfo{author}{\bibfnamefont{J.}~\bibnamefont{Nunn}}, \bibnamefont{and}
  \bibinfo{author}{\bibfnamefont{B.~J.} \bibnamefont{Sussman}},
  \bibinfo{journal}{J. Mod. Opt.} \textbf{\bibinfo{volume}{63}},
  \bibinfo{pages}{2005} (\bibinfo{year}{2016}).

\bibitem[{\citenamefont{Wang et~al.}(2021)\citenamefont{Wang, Luan, Qiao, Um,
  Zhang, Wang, Yuan, Gu, Zhang, and Kim}}]{wang2021single}
\bibinfo{author}{\bibfnamefont{P.}~\bibnamefont{Wang}},
  \bibinfo{author}{\bibfnamefont{C.-Y.} \bibnamefont{Luan}},
  \bibinfo{author}{\bibfnamefont{M.}~\bibnamefont{Qiao}},
  \bibinfo{author}{\bibfnamefont{M.}~\bibnamefont{Um}},
  \bibinfo{author}{\bibfnamefont{J.}~\bibnamefont{Zhang}},
  \bibinfo{author}{\bibfnamefont{Y.}~\bibnamefont{Wang}},
  \bibinfo{author}{\bibfnamefont{X.}~\bibnamefont{Yuan}},
  \bibinfo{author}{\bibfnamefont{M.}~\bibnamefont{Gu}},
  \bibinfo{author}{\bibfnamefont{J.}~\bibnamefont{Zhang}}, \bibnamefont{and}
  \bibinfo{author}{\bibfnamefont{K.}~\bibnamefont{Kim}}, \bibinfo{journal}{Nat.
  Commun.} \textbf{\bibinfo{volume}{12}}, \bibinfo{pages}{233}
  (\bibinfo{year}{2021}).

\bibitem[{\citenamefont{Lvovsky et~al.}(2009)\citenamefont{Lvovsky, Sanders,
  and Tittel}}]{lvovsky2009optical}
\bibinfo{author}{\bibfnamefont{A.~I.} \bibnamefont{Lvovsky}},
  \bibinfo{author}{\bibfnamefont{B.~C.} \bibnamefont{Sanders}},
  \bibnamefont{and} \bibinfo{author}{\bibfnamefont{W.}~\bibnamefont{Tittel}},
  \bibinfo{journal}{Nat. Photon.} \textbf{\bibinfo{volume}{3}},
  \bibinfo{pages}{706} (\bibinfo{year}{2009}).

\bibitem[{\citenamefont{Choi et~al.}(2010)\citenamefont{Choi, Goban, Papp,
  Van~Enk, and Kimble}}]{choi2010entanglement}
\bibinfo{author}{\bibfnamefont{K.}~\bibnamefont{Choi}},
  \bibinfo{author}{\bibfnamefont{A.}~\bibnamefont{Goban}},
  \bibinfo{author}{\bibfnamefont{S.}~\bibnamefont{Papp}},
  \bibinfo{author}{\bibfnamefont{S.~J.}~\bibnamefont{van~Enk}}, \bibnamefont{and}
  \bibinfo{author}{\bibfnamefont{H.}~\bibnamefont{Kimble}},
  \bibinfo{journal}{Nature (London)} \textbf{\bibinfo{volume}{468}},
  \bibinfo{pages}{412} (\bibinfo{year}{2010}).

\bibitem[{\citenamefont{Julsgaard et~al.}(2004)\citenamefont{Julsgaard,
  Sherson, Cirac, Fiur{\'a}{\v{s}}ek, and Polzik}}]{julsgaard2004experimental}
\bibinfo{author}{\bibfnamefont{B.}~\bibnamefont{Julsgaard}},
  \bibinfo{author}{\bibfnamefont{J.}~\bibnamefont{Sherson}},
  \bibinfo{author}{\bibfnamefont{J.~I.} \bibnamefont{Cirac}},
  \bibinfo{author}{\bibfnamefont{J.}~\bibnamefont{Fiur{\'a}{\v{s}}ek}},
  \bibnamefont{and} \bibinfo{author}{\bibfnamefont{E.~S.}
  \bibnamefont{Polzik}}, \bibinfo{journal}{Nature (London)}
  \textbf{\bibinfo{volume}{432}}, \bibinfo{pages}{482} (\bibinfo{year}{2004}).

\bibitem[{\citenamefont{Bradley et~al.}(2019)\citenamefont{Bradley, Randall,
  Abobeih, Berrevoets, Degen, Bakker, Markham, Twitchen, and
  Taminiau}}]{bradley2019ten}
\bibinfo{author}{\bibfnamefont{C.~E.} \bibnamefont{Bradley}},
  \bibinfo{author}{\bibfnamefont{J.}~\bibnamefont{Randall}},
  \bibinfo{author}{\bibfnamefont{M.~H.} \bibnamefont{Abobeih}},
  \bibinfo{author}{\bibfnamefont{R.}~\bibnamefont{Berrevoets}},
  \bibinfo{author}{\bibfnamefont{M.}~\bibnamefont{Degen}},
  \bibinfo{author}{\bibfnamefont{M.~A.} \bibnamefont{Bakker}},
  \bibinfo{author}{\bibfnamefont{M.}~\bibnamefont{Markham}},
  \bibinfo{author}{\bibfnamefont{D.}~\bibnamefont{Twitchen}}, \bibnamefont{and}
  \bibinfo{author}{\bibfnamefont{T.~H.} \bibnamefont{Taminiau}},
  \bibinfo{journal}{Phys. Rev. X} \textbf{\bibinfo{volume}{9}},
  \bibinfo{pages}{031045} (\bibinfo{year}{2019}).

\bibitem[{\citenamefont{Neumann et~al.}(2010)\citenamefont{Neumann, Kolesov,
  Naydenov, Beck, Rempp, Steiner, Jacques, Balasubramanian, Markham, Twitchen
  et~al.}}]{neumann2010quantum}
\bibinfo{author}{\bibfnamefont{P.}~\bibnamefont{Neumann}},
  \bibinfo{author}{\bibfnamefont{R.}~\bibnamefont{Kolesov}},
  \bibinfo{author}{\bibfnamefont{B.}~\bibnamefont{Naydenov}},
  \bibinfo{author}{\bibfnamefont{J.}~\bibnamefont{Beck}},
  \bibinfo{author}{\bibfnamefont{F.}~\bibnamefont{Rempp}},
  \bibinfo{author}{\bibfnamefont{M.}~\bibnamefont{Steiner}},
  \bibinfo{author}{\bibfnamefont{V.}~\bibnamefont{Jacques}},
  \bibinfo{author}{\bibfnamefont{G.}~\bibnamefont{Balasubramanian}},
  \bibinfo{author}{\bibfnamefont{M.}~\bibnamefont{Markham}},
  \bibinfo{author}{\bibfnamefont{D.}~\bibnamefont{Twitchen}},
  \bibnamefont{et~al.}, \bibinfo{journal}{Nat. Phys.}
  \textbf{\bibinfo{volume}{6}}, \bibinfo{pages}{249} (\bibinfo{year}{2010}).

\bibitem[{\citenamefont{Fuchs et~al.}(2011)\citenamefont{Fuchs, Burkard,
  Klimov, and Awschalom}}]{fuchs2011quantum}
\bibinfo{author}{\bibfnamefont{G.}~\bibnamefont{Fuchs}},
  \bibinfo{author}{\bibfnamefont{G.}~\bibnamefont{Burkard}},
  \bibinfo{author}{\bibfnamefont{P.}~\bibnamefont{Klimov}}, \bibnamefont{and}
  \bibinfo{author}{\bibfnamefont{D.}~\bibnamefont{Awschalom}},
  \bibinfo{journal}{Nat. Phys.} \textbf{\bibinfo{volume}{7}},
  \bibinfo{pages}{789} (\bibinfo{year}{2011}).

\bibitem[{\citenamefont{Ruskuc et~al.}(2022)\citenamefont{Ruskuc, Wu, Rochman,
  Choi, and Faraon}}]{ruskuc2022nuclear}
\bibinfo{author}{\bibfnamefont{A.}~\bibnamefont{Ruskuc}},
  \bibinfo{author}{\bibfnamefont{C.-J.} \bibnamefont{Wu}},
  \bibinfo{author}{\bibfnamefont{J.}~\bibnamefont{Rochman}},
  \bibinfo{author}{\bibfnamefont{J.}~\bibnamefont{Choi}}, \bibnamefont{and}
  \bibinfo{author}{\bibfnamefont{A.}~\bibnamefont{Faraon}},
  \bibinfo{journal}{Nature (London)} \textbf{\bibinfo{volume}{602}},
  \bibinfo{pages}{408} (\bibinfo{year}{2022}).

\bibitem[{\citenamefont{Hedges et~al.}(2010)\citenamefont{Hedges, Longdell, Li,
  and Sellars}}]{hedges2010efficient}
\bibinfo{author}{\bibfnamefont{M.~P.} \bibnamefont{Hedges}},
  \bibinfo{author}{\bibfnamefont{J.~J.} \bibnamefont{Longdell}},
  \bibinfo{author}{\bibfnamefont{Y.}~\bibnamefont{Li}}, \bibnamefont{and}
  \bibinfo{author}{\bibfnamefont{M.~J.} \bibnamefont{Sellars}},
  \bibinfo{journal}{Nature (London)} \textbf{\bibinfo{volume}{465}},
  \bibinfo{pages}{1052} (\bibinfo{year}{2010}).

\bibitem[{\citenamefont{Morton et~al.}(2008)\citenamefont{Morton, Tyryshkin,
  Brown, Shankar, Lovett, Ardavan, Schenkel, Haller, Ager, and
  Lyon}}]{morton2008solid}
\bibinfo{author}{\bibfnamefont{J.~J.} \bibnamefont{Morton}},
  \bibinfo{author}{\bibfnamefont{A.~M.} \bibnamefont{Tyryshkin}},
  \bibinfo{author}{\bibfnamefont{R.~M.} \bibnamefont{Brown}},
  \bibinfo{author}{\bibfnamefont{S.}~\bibnamefont{Shankar}},
  \bibinfo{author}{\bibfnamefont{B.~W.} \bibnamefont{Lovett}},
  \bibinfo{author}{\bibfnamefont{A.}~\bibnamefont{Ardavan}},
  \bibinfo{author}{\bibfnamefont{T.}~\bibnamefont{Schenkel}},
  \bibinfo{author}{\bibfnamefont{E.~E.} \bibnamefont{Haller}},
  \bibinfo{author}{\bibfnamefont{J.~W.} \bibnamefont{Ager}}, \bibnamefont{and}
  \bibinfo{author}{\bibfnamefont{S.}~\bibnamefont{Lyon}},
  \bibinfo{journal}{Nature (London)} \textbf{\bibinfo{volume}{455}},
  \bibinfo{pages}{1085} (\bibinfo{year}{2008}).

\bibitem[{\citenamefont{Petta et~al.}(2005)\citenamefont{Petta, Johnson,
  Taylor, Laird, Yacoby, Lukin, Marcus, Hanson, and
  Gossard}}]{petta2005coherent}
\bibinfo{author}{\bibfnamefont{J.~R.} \bibnamefont{Petta}},
  \bibinfo{author}{\bibfnamefont{A.~C.} \bibnamefont{Johnson}},
  \bibinfo{author}{\bibfnamefont{J.~M.} \bibnamefont{Taylor}},
  \bibinfo{author}{\bibfnamefont{E.~A.} \bibnamefont{Laird}},
  \bibinfo{author}{\bibfnamefont{A.}~\bibnamefont{Yacoby}},
  \bibinfo{author}{\bibfnamefont{M.~D.} \bibnamefont{Lukin}},
  \bibinfo{author}{\bibfnamefont{C.~M.} \bibnamefont{Marcus}},
  \bibinfo{author}{\bibfnamefont{M.~P.} \bibnamefont{Hanson}},
  \bibnamefont{and} \bibinfo{author}{\bibfnamefont{A.~C.}
  \bibnamefont{Gossard}}, \bibinfo{journal}{Science}
  \textbf{\bibinfo{volume}{309}}, \bibinfo{pages}{2180} (\bibinfo{year}{2005}).

\bibitem[{\citenamefont{Koppens et~al.}(2005)\citenamefont{Koppens, Folk,
  Elzerman, Hanson, Van~Beveren, Vink, Tranitz, Wegscheider, Kouwenhoven, and
  Vandersypen}}]{koppens2005control}
\bibinfo{author}{\bibfnamefont{F.~H.} \bibnamefont{Koppens}},
  \bibinfo{author}{\bibfnamefont{J.~A.} \bibnamefont{Folk}},
  \bibinfo{author}{\bibfnamefont{J.~M.} \bibnamefont{Elzerman}},
  \bibinfo{author}{\bibfnamefont{R.}~\bibnamefont{Hanson}},
  \bibinfo{author}{\bibfnamefont{L.~W.} \bibnamefont{Van~Beveren}},
  \bibinfo{author}{\bibfnamefont{I.~T.} \bibnamefont{Vink}},
  \bibinfo{author}{\bibfnamefont{H.-P.} \bibnamefont{Tranitz}},
  \bibinfo{author}{\bibfnamefont{W.}~\bibnamefont{Wegscheider}},
  \bibinfo{author}{\bibfnamefont{L.~P.} \bibnamefont{Kouwenhoven}},
  \bibnamefont{and} \bibinfo{author}{\bibfnamefont{L.~M.}
  \bibnamefont{Vandersypen}}, \bibinfo{journal}{Science}
  \textbf{\bibinfo{volume}{309}}, \bibinfo{pages}{1346} (\bibinfo{year}{2005}).

\bibitem[{\citenamefont{Taylor et~al.}(2003{\natexlab{a}})\citenamefont{Taylor,
  Marcus, and Lukin}}]{taylor2003long}
\bibinfo{author}{\bibfnamefont{J.}~\bibnamefont{Taylor}},
  \bibinfo{author}{\bibfnamefont{C.}~\bibnamefont{Marcus}}, \bibnamefont{and}
  \bibinfo{author}{\bibfnamefont{M.}~\bibnamefont{Lukin}},
  \bibinfo{journal}{Phys. Rev. Lett.} \textbf{\bibinfo{volume}{90}},
  \bibinfo{pages}{206803} (\bibinfo{year}{2003}{\natexlab{a}}).

\bibitem[{\citenamefont{Dobrovitski et~al.}(2006)\citenamefont{Dobrovitski,
  Taylor, and Lukin}}]{dobrovitski2006long}
\bibinfo{author}{\bibfnamefont{V.}~\bibnamefont{Dobrovitski}},
  \bibinfo{author}{\bibfnamefont{J.}~\bibnamefont{Taylor}}, \bibnamefont{and}
  \bibinfo{author}{\bibfnamefont{M.}~\bibnamefont{Lukin}},
  \bibinfo{journal}{Phys. Rev. B} \textbf{\bibinfo{volume}{73}},
  \bibinfo{pages}{245318} (\bibinfo{year}{2006}).

\bibitem[{\citenamefont{Ding et~al.}(2014)\citenamefont{Ding, Shi, You, and
  Zhang}}]{ding2014high}
\bibinfo{author}{\bibfnamefont{W.}~\bibnamefont{Ding}},
  \bibinfo{author}{\bibfnamefont{A.}~\bibnamefont{Shi}},
  \bibinfo{author}{\bibfnamefont{J.}~\bibnamefont{You}}, \bibnamefont{and}
  \bibinfo{author}{\bibfnamefont{W.}~\bibnamefont{Zhang}},
  \bibinfo{journal}{Phys. Rev. B} \textbf{\bibinfo{volume}{90}},
  \bibinfo{pages}{235421} (\bibinfo{year}{2014}).

\bibitem[{\citenamefont{Gangloff et~al.}(2019)\citenamefont{Gangloff,
  Ethier-Majcher, Lang, Denning, Bodey, Jackson, Clarke, Hugues, Le~Gall, and
  Atat{\"u}re}}]{gangloff2019quantum}
\bibinfo{author}{\bibfnamefont{D.}~\bibnamefont{Gangloff}},
  \bibinfo{author}{\bibfnamefont{G.}~\bibnamefont{{\'E}thier-Majcher}},
  \bibinfo{author}{\bibfnamefont{C.}~\bibnamefont{Lang}},
  \bibinfo{author}{\bibfnamefont{E.}~\bibnamefont{Denning}},
  \bibinfo{author}{\bibfnamefont{J.}~\bibnamefont{Bodey}},
  \bibinfo{author}{\bibfnamefont{D.}~\bibnamefont{Jackson}},
  \bibinfo{author}{\bibfnamefont{E.}~\bibnamefont{Clarke}},
  \bibinfo{author}{\bibfnamefont{M.}~\bibnamefont{Hugues}},
  \bibinfo{author}{\bibfnamefont{C.}~\bibnamefont{Le~Gall}}, \bibnamefont{and}
  \bibinfo{author}{\bibfnamefont{M.}~\bibnamefont{Atat{\"u}re}},
  \bibinfo{journal}{Science} \textbf{\bibinfo{volume}{364}},
  \bibinfo{pages}{62} (\bibinfo{year}{2019}).

\bibitem[{\citenamefont{Chekhovich et~al.}(2020)\citenamefont{Chekhovich,
  da~Silva, and Rastelli}}]{chekhovich2020nuclear}
\bibinfo{author}{\bibfnamefont{E.~A.} \bibnamefont{Chekhovich}},
  \bibinfo{author}{\bibfnamefont{S.~F.~C.} \bibnamefont{da~Silva}},
  \bibnamefont{and} \bibinfo{author}{\bibfnamefont{A.}~\bibnamefont{Rastelli}},
  \bibinfo{journal}{Nat. Nanotechnol.} \textbf{\bibinfo{volume}{15}},
  \bibinfo{pages}{999} (\bibinfo{year}{2020}).

\bibitem[{\citenamefont{Jackson et~al.}(2021)\citenamefont{Jackson, Gangloff,
  Bodey, Zaporski, Bachorz, Clarke, Hugues, Le~Gall, and
  Atat{\"u}re}}]{jackson2021quantum}
\bibinfo{author}{\bibfnamefont{D.~M.} \bibnamefont{Jackson}},
  \bibinfo{author}{\bibfnamefont{D.~A.} \bibnamefont{Gangloff}},
  \bibinfo{author}{\bibfnamefont{J.~H.} \bibnamefont{Bodey}},
  \bibinfo{author}{\bibfnamefont{L.}~\bibnamefont{Zaporski}},
  \bibinfo{author}{\bibfnamefont{C.}~\bibnamefont{Bachorz}},
  \bibinfo{author}{\bibfnamefont{E.}~\bibnamefont{Clarke}},
  \bibinfo{author}{\bibfnamefont{M.}~\bibnamefont{Hugues}},
  \bibinfo{author}{\bibfnamefont{C.}~\bibnamefont{Le~Gall}}, \bibnamefont{and}
  \bibinfo{author}{\bibfnamefont{M.}~\bibnamefont{Atat{\"u}re}},
  \bibinfo{journal}{Nat. Phys.} \textbf{\bibinfo{volume}{17}},
  \bibinfo{pages}{585} (\bibinfo{year}{2021}).

\bibitem[{\citenamefont{Gillard et~al.}(2022)\citenamefont{Gillard, Clarke, and
  Chekhovich}}]{gillard2022harnessing}
\bibinfo{author}{\bibfnamefont{G.}~\bibnamefont{Gillard}},
  \bibinfo{author}{\bibfnamefont{E.}~\bibnamefont{Clarke}}, \bibnamefont{and}
  \bibinfo{author}{\bibfnamefont{E.~A.} \bibnamefont{Chekhovich}},
  \bibinfo{journal}{Nat. Commun.} \textbf{\bibinfo{volume}{13}},
  \bibinfo{pages}{4048} (\bibinfo{year}{2022}).

\bibitem[{\citenamefont{W{\"u}st et~al.}(2016)\citenamefont{W{\"u}st, Munsch,
  Maier, Kuhlmann, Ludwig, Wieck, Loss, Poggio, and Warburton}}]{wust2016role}
\bibinfo{author}{\bibfnamefont{G.}~\bibnamefont{W{\"u}st}},
  \bibinfo{author}{\bibfnamefont{M.}~\bibnamefont{Munsch}},
  \bibinfo{author}{\bibfnamefont{F.}~\bibnamefont{Maier}},
  \bibinfo{author}{\bibfnamefont{A.~V.} \bibnamefont{Kuhlmann}},
  \bibinfo{author}{\bibfnamefont{A.}~\bibnamefont{Ludwig}},
  \bibinfo{author}{\bibfnamefont{A.~D.} \bibnamefont{Wieck}},
  \bibinfo{author}{\bibfnamefont{D.}~\bibnamefont{Loss}},
  \bibinfo{author}{\bibfnamefont{M.}~\bibnamefont{Poggio}}, \bibnamefont{and}
  \bibinfo{author}{\bibfnamefont{R.~J.} \bibnamefont{Warburton}},
  \bibinfo{journal}{Nat. Nanotechnol.} \textbf{\bibinfo{volume}{11}},
  \bibinfo{pages}{885} (\bibinfo{year}{2016}).

\bibitem[{\citenamefont{Chekhovich et~al.}(2015)\citenamefont{Chekhovich,
  Hopkinson, Skolnick, and Tartakovskii}}]{chekhovich2015suppression}
\bibinfo{author}{\bibfnamefont{E.}~\bibnamefont{Chekhovich}},
  \bibinfo{author}{\bibfnamefont{M.}~\bibnamefont{Hopkinson}},
  \bibinfo{author}{\bibfnamefont{M.}~\bibnamefont{Skolnick}}, \bibnamefont{and}
  \bibinfo{author}{\bibfnamefont{A.}~\bibnamefont{Tartakovskii}},
  \bibinfo{journal}{Nat. Commun.} \textbf{\bibinfo{volume}{6}},
  \bibinfo{pages}{6348} (\bibinfo{year}{2015}).

\bibitem[{\citenamefont{Chekhovich et~al.}(2017)\citenamefont{Chekhovich,
  Ulhaq, Zallo, Ding, Schmidt, and Skolnick}}]{chekhovich2017measurement}
\bibinfo{author}{\bibfnamefont{E.}~\bibnamefont{Chekhovich}},
  \bibinfo{author}{\bibfnamefont{A.}~\bibnamefont{Ulhaq}},
  \bibinfo{author}{\bibfnamefont{E.}~\bibnamefont{Zallo}},
  \bibinfo{author}{\bibfnamefont{F.}~\bibnamefont{Ding}},
  \bibinfo{author}{\bibfnamefont{O.}~\bibnamefont{Schmidt}}, \bibnamefont{and}
  \bibinfo{author}{\bibfnamefont{M.}~\bibnamefont{Skolnick}},
  \bibinfo{journal}{Nat. Mater.} \textbf{\bibinfo{volume}{16}},
  \bibinfo{pages}{982} (\bibinfo{year}{2017}).

\bibitem[{\citenamefont{{\'E}thier-Majcher
  et~al.}(2017)\citenamefont{{\'E}thier-Majcher, Gangloff, Stockill, Clarke,
  Hugues, Le~Gall, and Atat{\"u}re}}]{ethier2017improving}
\bibinfo{author}{\bibfnamefont{G.}~\bibnamefont{{\'E}thier-Majcher}},
  \bibinfo{author}{\bibfnamefont{D.}~\bibnamefont{Gangloff}},
  \bibinfo{author}{\bibfnamefont{R.}~\bibnamefont{Stockill}},
  \bibinfo{author}{\bibfnamefont{E.}~\bibnamefont{Clarke}},
  \bibinfo{author}{\bibfnamefont{M.}~\bibnamefont{Hugues}},
  \bibinfo{author}{\bibfnamefont{C.}~\bibnamefont{Le~Gall}}, \bibnamefont{and}
  \bibinfo{author}{\bibfnamefont{M.}~\bibnamefont{Atat{\"u}re}},
  \bibinfo{journal}{Phys. Rev. Lett.} \textbf{\bibinfo{volume}{119}},
  \bibinfo{pages}{130503} (\bibinfo{year}{2017}).

\bibitem[{\citenamefont{Reilly et~al.}(2008)\citenamefont{Reilly, Taylor,
  Petta, Marcus, Hanson, and Gossard}}]{reilly2008suppressing}
\bibinfo{author}{\bibfnamefont{D.}~\bibnamefont{Reilly}},
  \bibinfo{author}{\bibfnamefont{J.}~\bibnamefont{Taylor}},
  \bibinfo{author}{\bibfnamefont{J.}~\bibnamefont{Petta}},
  \bibinfo{author}{\bibfnamefont{C.}~\bibnamefont{Marcus}},
  \bibinfo{author}{\bibfnamefont{M.}~\bibnamefont{Hanson}}, \bibnamefont{and}
  \bibinfo{author}{\bibfnamefont{A.}~\bibnamefont{Gossard}},
  \bibinfo{journal}{Science} \textbf{\bibinfo{volume}{321}},
  \bibinfo{pages}{817} (\bibinfo{year}{2008}).

\bibitem[{\citenamefont{Evers et~al.}(2021)\citenamefont{Evers, Kopteva,
  Yugova, Yakovlev, Reuter, Wieck, Bayer, and Greilich}}]{evers2021suppression}
\bibinfo{author}{\bibfnamefont{E.}~\bibnamefont{Evers}},
  \bibinfo{author}{\bibfnamefont{N.}~\bibnamefont{Kopteva}},
  \bibinfo{author}{\bibfnamefont{I.}~\bibnamefont{Yugova}},
  \bibinfo{author}{\bibfnamefont{D.}~\bibnamefont{Yakovlev}},
  \bibinfo{author}{\bibfnamefont{D.}~\bibnamefont{Reuter}},
  \bibinfo{author}{\bibfnamefont{A.}~\bibnamefont{Wieck}},
  \bibinfo{author}{\bibfnamefont{M.}~\bibnamefont{Bayer}}, \bibnamefont{and}
  \bibinfo{author}{\bibfnamefont{A.}~\bibnamefont{Greilich}},
  \bibinfo{journal}{npj Quantum Inf.} \textbf{\bibinfo{volume}{7}},
  \bibinfo{pages}{60} (\bibinfo{year}{2021}).

\bibitem[{\citenamefont{Wu et~al.}(2016)\citenamefont{Wu, Ding, Shi, and
  Zhang}}]{wu2016inhomogeneous}
\bibinfo{author}{\bibfnamefont{N.}~\bibnamefont{Wu}},
  \bibinfo{author}{\bibfnamefont{W.}~\bibnamefont{Ding}},
  \bibinfo{author}{\bibfnamefont{A.}~\bibnamefont{Shi}}, \bibnamefont{and}
  \bibinfo{author}{\bibfnamefont{W.}~\bibnamefont{Zhang}},
  \bibinfo{journal}{Phys. Lett. A} \textbf{\bibinfo{volume}{380}},
  \bibinfo{pages}{2706} (\bibinfo{year}{2016}).

\bibitem[{\citenamefont{Denning et~al.}(2019)\citenamefont{Denning, Gangloff,
  Atat{\"u}re, M{\o}rk, and Le~Gall}}]{denning2019collective}
\bibinfo{author}{\bibfnamefont{E.~V.} \bibnamefont{Denning}},
  \bibinfo{author}{\bibfnamefont{D.~A.} \bibnamefont{Gangloff}},
  \bibinfo{author}{\bibfnamefont{M.}~\bibnamefont{Atat{\"u}re}},
  \bibinfo{author}{\bibfnamefont{J.}~\bibnamefont{M{\o}rk}}, \bibnamefont{and}
  \bibinfo{author}{\bibfnamefont{C.}~\bibnamefont{Le~Gall}},
  \bibinfo{journal}{Phys. Rev. Lett.} \textbf{\bibinfo{volume}{123}},
  \bibinfo{pages}{140502} (\bibinfo{year}{2019}).

\bibitem[{\citenamefont{Khaetskii et~al.}(2002)\citenamefont{Khaetskii, Loss,
  and Glazman}}]{khaetskii2002electron}
\bibinfo{author}{\bibfnamefont{A.~V.} \bibnamefont{Khaetskii}},
  \bibinfo{author}{\bibfnamefont{D.}~\bibnamefont{Loss}}, \bibnamefont{and}
  \bibinfo{author}{\bibfnamefont{L.}~\bibnamefont{Glazman}},
  \bibinfo{journal}{Phys. Rev. Lett.} \textbf{\bibinfo{volume}{88}},
  \bibinfo{pages}{186802} (\bibinfo{year}{2002}).

\bibitem[{\citenamefont{Merkulov et~al.}(2002)\citenamefont{Merkulov, Efros,
  and Rosen}}]{merkulov2002electron}
\bibinfo{author}{\bibfnamefont{I.~A.} \bibnamefont{Merkulov}},
  \bibinfo{author}{\bibfnamefont{A.~L.} \bibnamefont{Efros}}, \bibnamefont{and}
  \bibinfo{author}{\bibfnamefont{M.}~\bibnamefont{Rosen}},
  \bibinfo{journal}{Phys. Rev. B} \textbf{\bibinfo{volume}{65}},
  \bibinfo{pages}{205309} (\bibinfo{year}{2002}).

\bibitem[{\citenamefont{Philippopoulos
  et~al.}(2020)\citenamefont{Philippopoulos, Chesi, and
  Coish}}]{philippopoulos2020first}
\bibinfo{author}{\bibfnamefont{P.}~\bibnamefont{Philippopoulos}},
  \bibinfo{author}{\bibfnamefont{S.}~\bibnamefont{Chesi}}, \bibnamefont{and}
  \bibinfo{author}{\bibfnamefont{W.}~\bibnamefont{Coish}},
  \bibinfo{journal}{Phys. Rev. B} \textbf{\bibinfo{volume}{101}},
  \bibinfo{pages}{115302} (\bibinfo{year}{2020}).

\bibitem[{\citenamefont{Slichter}(1990)}]{slichter2013principles}
\bibinfo{author}{\bibfnamefont{C.~P.} \bibnamefont{Slichter}},
  \emph{\bibinfo{title}{Principles of magnetic resonance}}
  (\bibinfo{publisher}{Springer-Verlag}, \bibinfo{address}{Berlin},
  \bibinfo{year}{1990}), \bibinfo{edition}{3rd} ed.

\bibitem[{\citenamefont{Johnson et~al.}(2005)\citenamefont{Johnson, Petta,
  Taylor, Yacoby, Lukin, Marcus, Hanson, and Gossard}}]{johnson2005triplet}
\bibinfo{author}{\bibfnamefont{A.}~\bibnamefont{Johnson}},
  \bibinfo{author}{\bibfnamefont{J.}~\bibnamefont{Petta}},
  \bibinfo{author}{\bibfnamefont{J.}~\bibnamefont{Taylor}},
  \bibinfo{author}{\bibfnamefont{A.}~\bibnamefont{Yacoby}},
  \bibinfo{author}{\bibfnamefont{M.}~\bibnamefont{Lukin}},
  \bibinfo{author}{\bibfnamefont{C.}~\bibnamefont{Marcus}},
  \bibinfo{author}{\bibfnamefont{M.}~\bibnamefont{Hanson}}, \bibnamefont{and}
  \bibinfo{author}{\bibfnamefont{A.}~\bibnamefont{Gossard}},
  \bibinfo{journal}{Nature (London)} \textbf{\bibinfo{volume}{435}},
  \bibinfo{pages}{925} (\bibinfo{year}{2005}).

\bibitem[{\citenamefont{Taylor et~al.}(2003{\natexlab{b}})\citenamefont{Taylor,
  Imamoglu, and Lukin}}]{taylor2003controlling}
\bibinfo{author}{\bibfnamefont{J.}~\bibnamefont{Taylor}},
  \bibinfo{author}{\bibfnamefont{A.}~\bibnamefont{Imamoglu}}, \bibnamefont{and}
  \bibinfo{author}{\bibfnamefont{M.}~\bibnamefont{Lukin}},
  \bibinfo{journal}{Phys. Rev. Lett.} \textbf{\bibinfo{volume}{91}},
  \bibinfo{pages}{246802} (\bibinfo{year}{2003}{\natexlab{b}}).

\bibitem[{\citenamefont{Choi et~al.}(2020)\citenamefont{Choi, Zhou, Knowles,
  Landig, Choi, and Lukin}}]{choi2020robust}
\bibinfo{author}{\bibfnamefont{J.}~\bibnamefont{Choi}},
  \bibinfo{author}{\bibfnamefont{H.}~\bibnamefont{Zhou}},
  \bibinfo{author}{\bibfnamefont{H.~S.} \bibnamefont{Knowles}},
  \bibinfo{author}{\bibfnamefont{R.}~\bibnamefont{Landig}},
  \bibinfo{author}{\bibfnamefont{S.}~\bibnamefont{Choi}}, \bibnamefont{and}
  \bibinfo{author}{\bibfnamefont{M.~D.} \bibnamefont{Lukin}},
  \bibinfo{journal}{Phys. Rev. X} \textbf{\bibinfo{volume}{10}},
  \bibinfo{pages}{031002} (\bibinfo{year}{2020}).

\bibitem[{\citenamefont{Coish and Loss}(2004)}]{coish2004hyperfine}
\bibinfo{author}{\bibfnamefont{W.}~\bibnamefont{Coish}} \bibnamefont{and}
  \bibinfo{author}{\bibfnamefont{D.}~\bibnamefont{Loss}},
  \bibinfo{journal}{Phys. Rev. B} \textbf{\bibinfo{volume}{70}},
  \bibinfo{pages}{195340} (\bibinfo{year}{2004}).

\bibitem[{\citenamefont{Bortz and Stolze}(2007)}]{bortz2007exact}
\bibinfo{author}{\bibfnamefont{M.}~\bibnamefont{Bortz}} \bibnamefont{and}
  \bibinfo{author}{\bibfnamefont{J.}~\bibnamefont{Stolze}},
  \bibinfo{journal}{Phys. Rev. B} \textbf{\bibinfo{volume}{76}},
  \bibinfo{pages}{014304} (\bibinfo{year}{2007}).

\bibitem[{\citenamefont{Gaudin}(1976)}]{gaudin1976diagonalisation}
\bibinfo{author}{\bibfnamefont{M.}~\bibnamefont{Gaudin}}, \bibinfo{journal}{J.
  Phys. (France)} \textbf{\bibinfo{volume}{37}}, \bibinfo{pages}{1087}
  (\bibinfo{year}{1976}).

\bibitem[{\citenamefont{Faribault and
  Schuricht}(2013)}]{faribault2013integrability}
\bibinfo{author}{\bibfnamefont{A.}~\bibnamefont{Faribault}} \bibnamefont{and}
  \bibinfo{author}{\bibfnamefont{D.}~\bibnamefont{Schuricht}},
  \bibinfo{journal}{Phys. Rev. Lett.} \textbf{\bibinfo{volume}{110}},
  \bibinfo{pages}{040405} (\bibinfo{year}{2013}).

\bibitem[{\citenamefont{Schliemann et~al.}(2003)\citenamefont{Schliemann,
  Khaetskii, and Loss}}]{schliemann2003electron}
\bibinfo{author}{\bibfnamefont{J.}~\bibnamefont{Schliemann}},
  \bibinfo{author}{\bibfnamefont{A.}~\bibnamefont{Khaetskii}},
  \bibnamefont{and} \bibinfo{author}{\bibfnamefont{D.}~\bibnamefont{Loss}},
  \bibinfo{journal}{J. Phys.: Condens. Matter} \textbf{\bibinfo{volume}{15}},
  \bibinfo{pages}{R1809} (\bibinfo{year}{2003}).

\bibitem[{\citenamefont{Dobrovitski and
  De~Raedt}(2003)}]{dobrovitski2003efficient}
\bibinfo{author}{\bibfnamefont{V.}~\bibnamefont{Dobrovitski}} \bibnamefont{and}
  \bibinfo{author}{\bibfnamefont{H.}~\bibnamefont{De~Raedt}},
  \bibinfo{journal}{Phys. Rev. E} \textbf{\bibinfo{volume}{67}},
  \bibinfo{pages}{056702} (\bibinfo{year}{2003}).

\bibitem[{\citenamefont{Zhang et~al.}(2010)\citenamefont{Zhang, Hu, Zhuang,
  You, and Liu}}]{zhang2010protection}
\bibinfo{author}{\bibfnamefont{W.}~\bibnamefont{Zhang}},
  \bibinfo{author}{\bibfnamefont{J.-L.} \bibnamefont{Hu}},
  \bibinfo{author}{\bibfnamefont{J.}~\bibnamefont{Zhuang}},
  \bibinfo{author}{\bibfnamefont{J.}~\bibnamefont{You}}, \bibnamefont{and}
  \bibinfo{author}{\bibfnamefont{R.-B.} \bibnamefont{Liu}},
  \bibinfo{journal}{Phys. Rev. B} \textbf{\bibinfo{volume}{82}},
  \bibinfo{pages}{045314} (\bibinfo{year}{2010}).

\end{thebibliography}
\end{document}